\documentclass[crop]{CSL}%%%%where CSL is the template name

\jname{Global Sustainability}%

\usepackage{graphics} 
\usepackage{anyfontsize}
\usepackage[utf8]{inputenc}
\begin{document}

\supertitle{Intelligence briefing}

%\title[recto running head]{How to use the CSL \LaTeX\ class}
%\title{Can Antarctica tip climate policy? \\
\title{Domino Effects in the Earth System -- The potential role of wanted tipping points}
%-- Anticipated Climate Impacts and rapid social change}
%Commentary: Can anticipated climate impacts induce social change? }

%\author[verso running head]{First author$^{1}$, Second author$^{2}$, Third author$^{2}$, Fourth author$^{2}$, Fifth author$^{2}$ and Seventh author$^{2}$}
\author[Wiedermann, Winkelmann et al.]{Marc Wiedermann$^{1*}$, Ricarda Winkelmann$^{1,2*}$, Jonathan F.
  Donges$^{1,3}$, Christina Eder$^{4}$,
Jobst Heitzig$^{1}$, Alexia Katsanidou$^{4,5}$, E. Keith Smith$^{4}$}

%\address{\add{1}{First author address} and \add{2}{Second author address}}
\address{
  \add{1}{Potsdam Institute for Climate Impact Research, Member of the Leibniz
    Association, Telegrafenberg A31, 14473 Potsdam, Germany}\\ \add{2}{Institute
    of Physics and Astronomy, University of Potsdam, Potsdam, Germany}\\
\add{3}{Stockholm Resilience Centre, Stockholm University, Kr\"aftriket
  2B, 114 19 Stockholm, Sweden}\\
\add{4}{GESIS -- Leibniz Institute for the Social Sciences, Unter Sachsenhausen 6-8, 50667 Cologne, Germany} \\
\add{5}{Institute of Sociology and Social Psychology, University of Cologne,
  Cologne, Germany} \\ \add{*}{shared first authorship}}

%\corres{\name{First author} \email{xxxx@xxxx.xxx.xx}}
\corres{\name{Ricarda Winkelmann} \email{ricarda.winkelmann@pik-potsdam.de} and
\name{Marc Wiedermann} \email{marc.wiedermann@pik-potsdam.de}}

%\begin{abstract}
%This sample is a guideline for preparing technical papers using \LaTeX\
%for CSL manuscript submission. It contains the documentation for CSL \LaTeX\
%class file, which implements the layout of the manuscript for CSL journal.
%This sample file uses a class file named \texttt{CSL.cls} where the authors
%should use during their manuscript preparation.
%\end{abstract}
\begin{abstract} % TODO MARC AND RICA
Vital parts of the climate system, such as the West Antarctic Ice Sheet, are at
risk even within the aspired aims of the Paris Agreement to limit global
temperature rise to $1.5$ -- $2^\circ$C. 
These so-called natural tipping elements are characterized by rapid qualitative shifts in their states once a critical threshold, e.g., of global mean temperature, is transgressed. We argue that the prevention of such \textit{unwanted} tipping through effective climate policies may  critically depend on cascading Domino effects from (anticipated) climate impacts to emission reductions via \textit{wanted} social tipping in attitudes, behaviors and policies. Specifically, the latter has recently been noted as a key aspect in addressing contemporary global challenges, such as climate change, via self-amplifying positive feedbacks similar to the processes observable in climate tipping elements. Our discussion is supported with data on commonly observed linkages between (subjective) climate change knowledge, concern and consequential potential action and we argue that the presence of such interrelations serves as a necessary condition for the possibility of rapid climate action.
\end{abstract}

%\begin{keypoints}
%\item Individuals can collectively act upon climate change and tip policies to
%  reach the Paris Agreement
%\item Tipping and collective action may be triggered by climate impacts and concern 
%\item Novel data showing a proposed tipping chain from knowledge to action
%via concern is introduced
%\end{keypoints}

%\selfcitation{Erhart C, Lutz S, Mutschler MA, Scharf PA, Walter T, Mantz H, Weigel R (2019). Compact polarimetry for automotive applications. International Journal of Microwave and Wireless Technologies 1--7. https://doi.org/10.1017/xxxxx}

%\received{xx xxxx xxxx}

%\revised{xx xxxx xxxx}

%\accepted{xx xxxx xxxx}

\maketitle

\Fpagebreak

\begin{figure*}[b!]
\centering
\includegraphics[width=.7\textwidth]{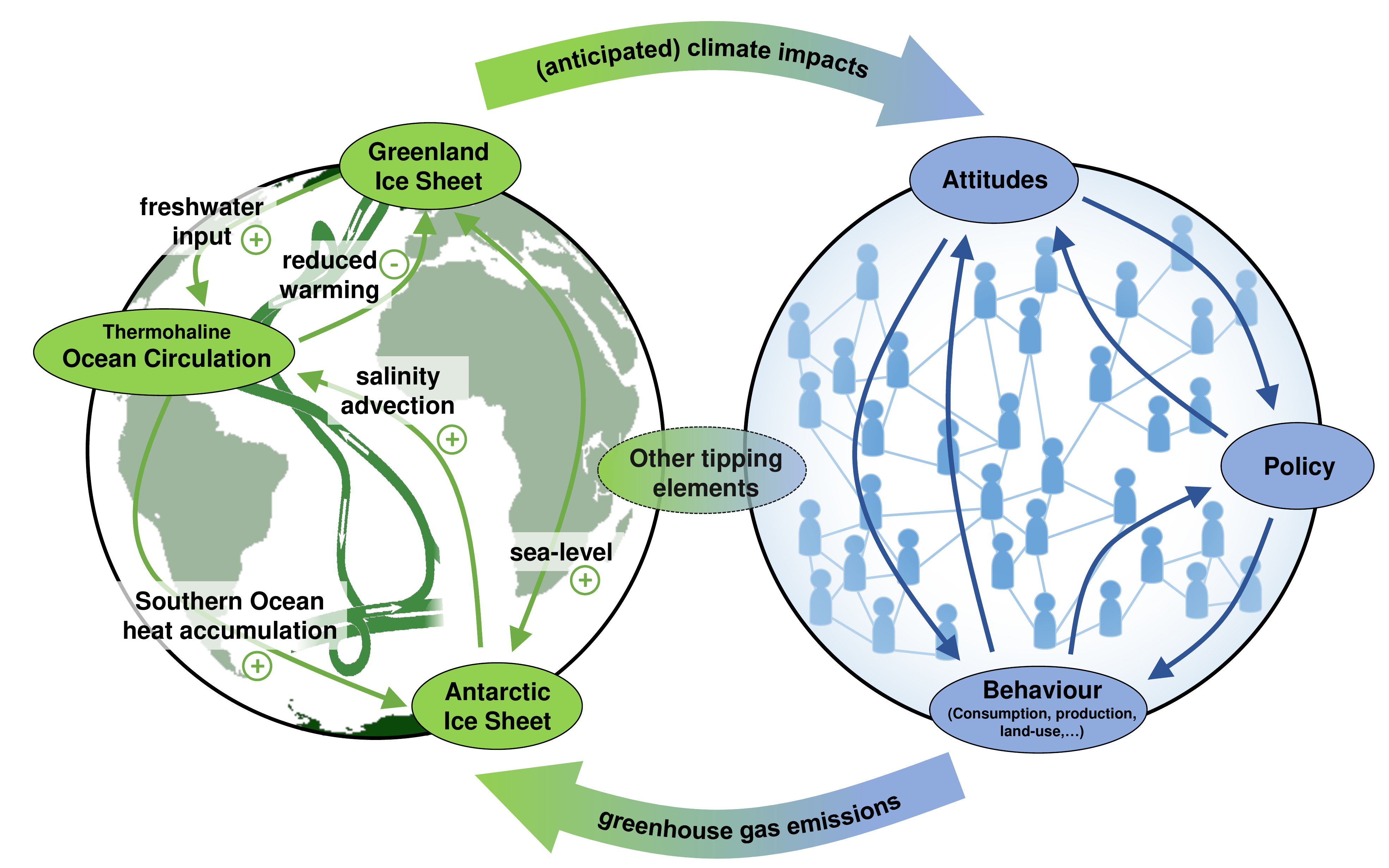}
\caption{Domino effects in the Earth System. The knowledge of (anticipated or already
observable) impacts caused by climate tipping elements (in green) such as the Antarctic Ice
Sheet, Greenland Ice Sheet and the Thermohaline Ocean Circulation could lead to social
tipping (in blue) via changes in attitudes, policies and -- ultimately -- behavior (for instance
changes in consumption, production or land-use). In turn, processes within the social sphere
can strongly affect the climate system through changes in greenhouse gas emissions.
Green arrows denote possible feedbacks between the climate tipping elements. Plus (minus)
signs indicate that the tipping of one element is likely to enhance (decrease) the possibility of
another. Blue arrows indicate possible interactions in the social sphere.}
\label{fig:figure1}
\end{figure*}

\section{How climate tipping became real -or- The unwanted tipping point}
There may have been a turning point in our thinking about climate change: In 2014,
several research groups suggested that the Amundsen region of West Antarctica
had entered a new dynamic regime, with glacier retreat controlled by the marine
ice-sheet instability, which could lead to its collapse in the long-term
\citep[e.g.,][]{rignot_widespread_2014}. 
Changes in the surrounding ocean circulation have
initiated internal ice dynamics that essentially cannot be reversed. They will
likely lead to a destabilization of this entire ice basin and subsequently to
global sea-level rise \citep{joughin_marine_2014,feldmann_collapse_2015}. A lot
is at stake: this part of West Antarctica alone holds enough ice to raise
global sea level by more than one meter \citep{bamber_reassessment_2009}, while all of the
marine ice basins in Antarctica together have a sea-level equivalent of almost
20 meters \citep{deconto_contribution_2016}.

The Antarctic Ice Sheet is only one of several crucial parts of the climate
system which --
though massive in size -- are so fragile that they can be severely disrupted by human activity.
For these so-called tipping elements, a small perturbation can suffice to push them into a
qualitatively different mode of operation \citep{schellnhuber_tipping_2009}. Several such large-scale
tipping elements have been identified in the climate system
\citep{lenton_tipping_2008,levermann_potential_2012}. Among the candidates with the most severe and far-reaching consequences for
the entire planet are the Thermohaline Ocean Circulation
\citep{caesar_observed_2018} and the
Greenland \citep{robinson_multistability_2012} and Antarctic ice sheets
\citep{winkelmann_combustion_2015} (left
part of Fig.~\ref{fig:figure1}). The risk of transgressing their critical thresholds increases with continued
greenhouse gas emissions and global warming \citep{church_sea_2013}. The ice-sheets, along
with coral reefs, mountain glaciers and the Arctic summer sea-ice, might in fact reach their
tipping point even within the Paris Agreement's target temperature range of 1.5 to 2$^\circ$ C of global warming 
 \citep{schellnhuber_why_2016}. Interactions
between these climate tipping elements  \citep{kriegler_imprecise_2009} could trigger cascading effects
and positive feedbacks that would further amplify global warming
 \citep{steffen_trajectories_2018,lenton_climate_2019} (see potential feedbacks in Fig.~\ref{fig:figure1}).

With the recent observations from the Amundsen basin
\citep{rignot_widespread_2014}, the rather
theoretical notion of climate tipping has become a reality. Declared a ``holy shit moment of
global warming'' by public media  \citep{mooney_this_2014}, the implications are being widely discussed in
science and the public  \citep{lenton_climate_2019}.

\section{The need for social tipping -or- The wanted tipping point}
\begin{figure*}[t!]
\centering
\includegraphics[width=.7\textwidth]{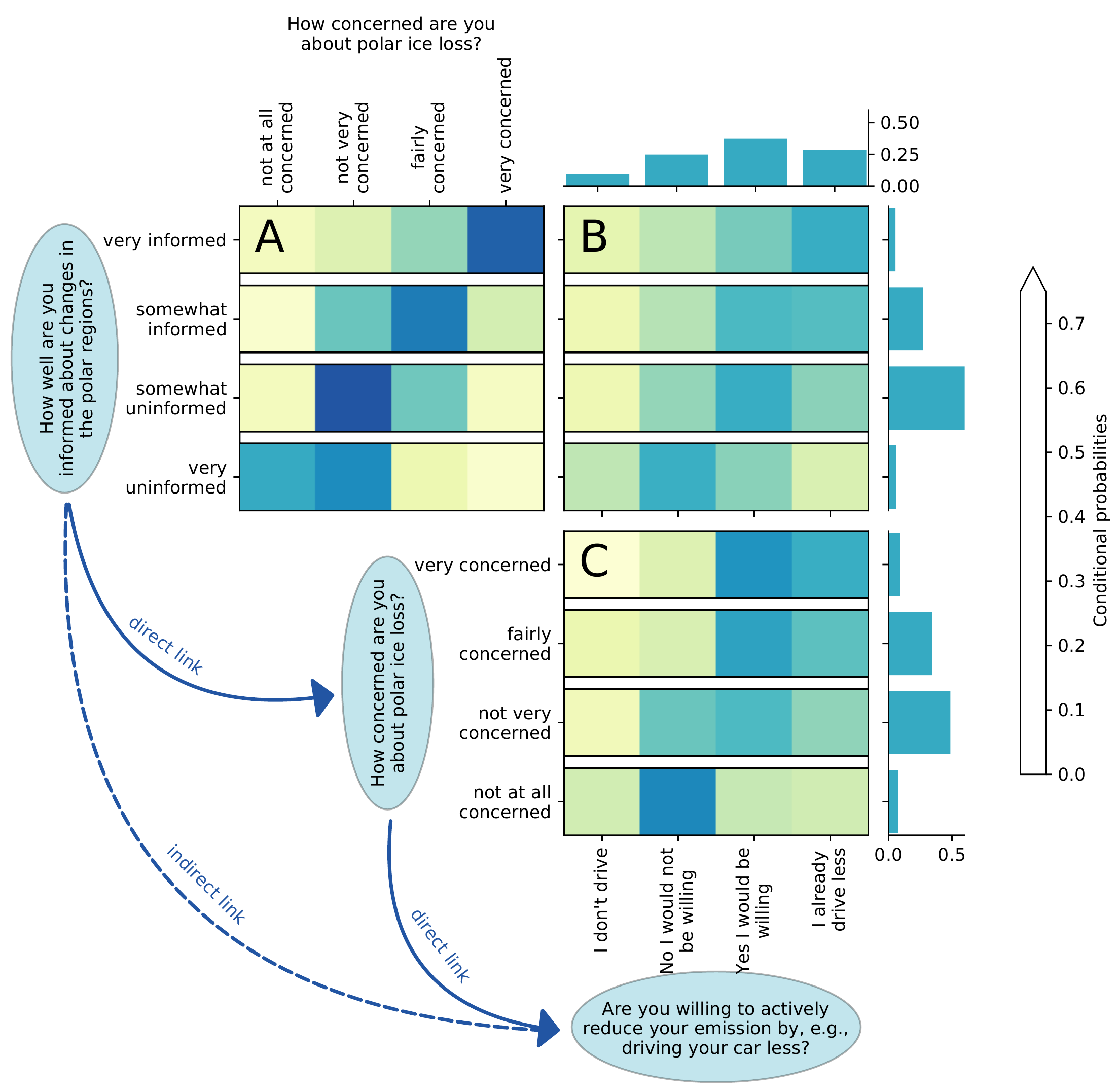}
\caption{Climate knowledge leads to action via concern. Results of a social survey from
Germany consisting of three questions regarding the loss of polar ice (see blue shaded
ellipses). The exact questions (in German, with translation into English) are given in the
Supplementary Information. The wording given in the labels presents a synthesis of the
original introductory and question texts. Colors in A,B,C indicate conditional probabilities for
answers on the horizontal axis given an answer on the vertical axis. The bar-charts indicate
the relative frequency of each given answer.}
\label{fig:figure2}
% We recommend that you provide the native width and height (natwidth, natheight) of your figures.
% Specifying native dimensions ensures that your figures are properly scaled
\end{figure*}

Without a rapid reduction in anthropogenic greenhouse gas emissions, sea-level rise from
the ice-sheets' mass loss will continue for centuries or even millennia, changing the face of
our planet as we know it  \citep{clark_consequences_2016}. In order to keep global warming
to ``well below
$2^\circ$C'' compared to preindustrial levels and thereby avoid some of the most severe climate change impacts on the long-term,
swift action is required: net-zero greenhouse gas emissions have to be achieved around mid-century, and the peak in emissions as early as 2020 \citep{figueres_three_2017}. This translates
into roughly halving CO\textsubscript{2} emissions each decade until 2050 in order to meet the Paris climate target with a 75\% chance \citep{rockstrom_roadmap_2017}. However, in reality, global emissions have
generally been rising over the past decades \citep{figueres_three_2017}.

We are thus facing a severe clash of time-scales, with the need to reverse the
current trend in emissions in just a few years, while climate impacts will
continue to manifest well beyond the end of this century. In order to overcome
this dilemma and to meet the ambitious goals of the Paris Agreement, the amplifying feedbacks between climate tipping elements need to be matched with self-reinforcing effects within social systems bringing about rapid societal changes, or \textit{social tipping}. This notion of social tipping, where a relatively small perturbation shifts a social system into a qualitatively different state via self-amplifying effects, has received recent scholarly interest  \citep{milkoreit_defining_2018, farmer_sensitive_2019, david_tabara_positive_2018}. As such, this definition of social tipping by means of positive feedback effects closely aligns with the proposed definitions of climate tipping elements \citep{lenton_tipping_2008,levermann_potential_2012}.

% TODO RICA: HERE, ADD 2 MORE SENTENCES ON SOCIAL TIPPING LIT
Societal transformation, where a system shifts from one state to a qualitatively different one has often been a key marking of
historical transformations. These shifts can be sudden, such as the ``Velvet
Revolution'' in communist Czechoslovakia, where the ruling party fell within a
matter of a few days, consequently leading to an end of the command economy and the establishing of a parliamentary republic. On the other hand, such shifts can also occur on comparably slower time-scale, like the decades-long ``civil rights
movement'' in the United States, culminating in the 1960's with a series of legislation banning formalized discriminatory practices, guaranteeing equal access to education, labor, housing and voting rights. In addition rapid societal change can even be caused by
unanticipated forces, such as the Fukushima Disaster in Japan affecting rapid
shifts away from nuclear energy production in Germany.

% ADD 1 sentence on why this is the "wanted" tipping point in contrast to the "unwanted" climate tipping point ambiguity / normative judgement (cite FARMER)
It remains to note that, in contrast to most natural tipping elements, social tipping elements can have positive or negative consequences for the coupled socio-ecological Earth system. As such, their classification, much more than for natural tipping elements, is often based on normative judgement \citep{david_tabara_positive_2018, farmer_sensitive_2019}. 

% TODO KEITH NEW PARAGRAPH Describe reason for looking at attitudes, behaviors and policies, as well as the relationships between them
\section{Anticipated climate impacts as potential drivers of social change}
It has been well established that meeting the aims of the Paris Agreement is not only
necessary but also still physically feasible in the climate system
\cite[e.g.,][]{schellnhuber_why_2016}. However, the questions remains as to what such feasibility entails in the social sphere.

Effective mitigation of climate change requires behavioral change on the individual level, but,
importantly, also the implementation of coordinated climate policies. Climate action on both
levels requires a sufficiently high level of awareness and climate change concern to allow
spreading of individual behavioral change on social networks and the support of global
climate policies through democratic institutions. A number of factors including values,
attitudes and political orientations, as well as other indicators, such as demographic
characteristics, knowledge, risk perception and trust, have all been linked to climate change
awareness and concern \citep{hornsey_meta-analyses_2016}. More importantly, subjective climate change knowledge, or climate literacy, has
been shown to be a strong positive predictor of such climate change concern
\citep{kahan_polarizing_2012}. At the same time, climate change concern has been shown to be strongly related to engagement in environmentally friendly actions, albeit less strongly than with climate policy
support \citep{hornsey_meta-analyses_2016}.

Given the need for rapid social changes, we question whether it is possible that anticipated impacts of climate change can affect behaviors and/or increase the likelihood of climate change ameliorative policies. Generally, the perceptions of risks have been broadly noted as key drivers of behavior, such that the more an individual perceives a certain risk, the more likely they are to act in ways that aim to minimize these risks. The role of such anticipated risks has been discussed with respect to a broad range of behaviors and aspects, such as health \citep{brewer_meta-analysis_2007,floyd_meta-analysis_2000}, financial
decision making \citep{weber_role_2004,weber_who_2012}, or sexual
behaviors \citep{benthin_psychometric_1993}. Specifically with respect to
climate change, increased risk perceptions have been found to motivate
individuals to engage in more climate  friendly
behaviors \citep{oconnor_risk_1999,semenza_public_2008} as well as support
climate policies \citep{leiserowitz_relationship_2013,smith_social_2018}.
%Further, subjective climate change knowledge or climate literacy have been shown to be a strong positive predictor of such climate change concern \citep{kahan_polarizing_2012}.

Drawing from environmental psychology, \cite{kollmuss_mind_2002} establish a theoretical framework 
noting the interplay between internal (e.g. motivation, values, attitudes) and external factors (e.g. institutional, economic, cultural) driving individual likelihood to engage in environmentally friendly behaviors. Internal motivations, such as concerns, can drive pro-environmental actions, most effectively when they are in synergy with other internal and external factors. And while knowledge of climate change is related to concerns, the relationship between knowledge and behavioral change is often indirect, filtered via external (institutional, economic, cultural) and internal (motivation) factors. Previous research also noted positive relationships between knowledge and concerns for the polar regions \citep{hamilton_who_2008,hamilton_did_2012}, but as of current, there is limited empirical research on the drivers of environmental actions in relation to the such aforementioned changes. 

To illustrate the potential of anticipated climate impacts to induce a change in people's attitude we present an exploratory analysis of the relationship between knowledge, concern and willingness to engage in
behavioral ameliorative action in response to observations in the polar regions, Fig.~\ref{fig:figure2}. Specifically we utilize a novel survey of 2904 Germany respondents in the German GESIS Panel (August to
October 2017) \citep{gesis_gesis_2018}, a nationally representative, probability-based, mixed
mode access panel, utilizing online and mail-in responses
\citep{bosnjak_establishing_2018}.

Broadly this data suggests that knowledge about changes in the polar regions is directly linked to concerns for this area, and further, concerns are positively related to the willingness to engage in ameliorative actions (in this case driving one's car less). Hence, we find the expected indirect link between (subjective) knowledge and one's willingness to change certain  behaviors that appears to be via the corresponding concerns. As such, there is preliminary evidence that concerns for the changes in the polar region may effect change in people's behaviors. These observed links between subjective knowledge and concern, as well as concern and potential  action are a necessary condition for behavioral and social transformation, i.e., social tipping to occur. In other words if either a direct link between knowledge and concern or concern and  behavior would be absent such transformations or tipping would possibly be far less likely to occur. 
The present analysis mainly intends to show that changes in  behavior may come about as the result of transformations along such a chain of processes. We show that the (German) public seems to recognize and anticipate changes in the polar region which they consequently see as a potential threat to future generations. Assessing whether this state ultimately leads to environmental action may be determined by a multitude of additional variables that require further in-depth analysis. 
%As such, our preliminary data shows that the current discourse on changes in the Polar regions at least provides a possible and favourable pathway along which the knowledge about global change has the potential to ultimately lead to changes in individual actions. 

Hence, further research needs to engage in understanding the possible mechanisms and potential drivers behind the relatinship between knowledge, concerns and behaviors.  As noted by  \cite{kollmuss_mind_2002}, several barriers limit the possibility for behavioral changes, such as higher costs of engaging in new behaviors \citep{diekmann_green_2003}, or limiting cultural, economic or political factors. We also note that knowledge may be further fostered by novel foci in education or increased media coverage. In addition, behavioral change can certainly also result from processes other than subjective knowledge and concern, e.g., new technologies or changing socio-economic conditions. Ultimately, individual actions themselves may be followed by emergent collective action as well as consequential environmental regulations and climate policies.

\section{Domino effects in the Earth System}
In summary, we ask, somewhat provocatively, whether changes in perceptions of natural dynamics in the Polar regions can raise concerns about climate change enough to tip behaviors and climate policies for achieving substantive reductions in greenhouse gas emissions \citep{shi_public_2015,kahan_polarizing_2012}. 

The answer seems to be two-fold: A priori, it may come as a surprise that people are concerned about
ongoing or future changes in Antarctica at all, since sea-level rise through ice loss --
compared to typical individual and political time horizons -- is a very slow process. The most
severe impacts will most likely only manifest beyond the lifetime of even our grandchildren’s
generation  \citep{clark_consequences_2016}. Therefore, it is all the more remarkable that knowledge about the dynamics in polar regions still correlates with an increased concern about climate change
in general, as indicated by the survey data. Ultimately, this concern can (as suggested by our
data as well as earlier studies in, e.g.,~\cite{hornsey_meta-analyses_2016}) have a positive impact on
attitudes towards climate change and the associated willingness to engage in
environmentally friendly actions. This clearly shows a potential for societal change as a
response to anticipated climate tipping. 

Further, increased concerns towards the environment have the potential to effect policy changes. Changes to public opinion can punctuate previously stable and ‘sticky’ institutions, leading to policy change \citep{baumgartner_agendas_2010}. Also, broad social activism can spur
generation new political coalitions, or effect a change shift in the priorities
of existing ones \citep{sabatier_advocacy_1988,weible_theories_2017}.

However, it would take a number of
steps, i.e., policy and/or economic changes to have a direct impact on emissions that in turn
reduce the risk for a tipping of, e.g, the Antarctic Ice Sheet. This implies that social tipping as
a whole is more than the change in a single observable, but rather a possible cascading
domino effect of intertwined processes, e.g., through social, political and financial networks.

% TODO MARC: ADD PARAGRAPH ON FUTURE WORK
Future work should thoroughly address several issues that so far hinder an in-depth analysis of social tipping elements in the Earth System. First, a concise definition of social tipping should be developed as the basis for community-driven efforts to classify and investigate the dynamics behind past and possible future tipping events. We suggest that such an endeavour requires a broad discourse among an interdiscplinary community of social and natural scientists with the goal to develop a common terminology and understanding of sudden shifts in social systems. The process that lead to a proposal for a definition of natural tipping elements \citep{lenton_tipping_2008} can serve as a guideline to develop similar concepts for social tipping as well. From there, theoretical and numerical models of social tipping should be developed to foster the understanding of underlying (individual) micro- and (emerging) macrodynamics that eventually lead to the emergence of social tipping points. Such models should be validated against (longitudinal) social science data to test for their validity, better inform the estimation of involved parameters and to prioritize the inclusion of underlying dynamics and processes. With respect to the data presented above, future work should aim at obtaining longitudinal observations of knowledge, concern and associated action or  behavior to better understand changes in the temporal feedbacks between the observed variables. Such longer running data would, together with further treatments of potential biases, allow for statements other than exemplary providing a current snapshot of public discourses and possible pathways for social transformation. Specifically, such an assessment would expand our understanding of the complex interplay between knowledge, concern and action beyond the previously observed direct and indirect patterns that are recaptured by our analysis.

Hence, it requires joint efforts from different disciplines, such as the social sciences, physical Earth system sciences as well as mathematics and numerics to investigate the processes behind potential Domino Effects in the intertwined Earth System where \textit{wanted} social tipping has the potential to prevent crossing dangerous \textit{unwanted} climate tipping points.

\ack[Acknowledgement]{
We are very grateful to John Schellnhuber, Wolfgang Lucht,
Doyne Farmer and Tim Lenton for their helpful comments and inspiring discussions.
The data presented in this work is accessible at
\url{https://www.gesis.org/gesis-panel/data/}.
}
\\\\
\noindent\textbf{Conflict of Interest Statement:}
The authors declare no conflict of
interest.
\\\\
\noindent\textbf{Author Contributions:} MW, RW, and JFD designed the study. MW
analysed the data. MW and RW wrote the article. JFD and JH contributed to the
writing of the article. CE, AK and EKS contributed to the writing of the
article and contributed to the panel survey. RW supervised the study. All authors
developed the panel questions.
\\\\
\noindent\textbf{Financial Support:} 
This work has been supported by the Leibniz Association (project DOMINOES), the European Research Council project Earth Resilience in the Anthropocene (743080 ERA), the
Stordalen Foundation via the Planetary Boundary Research Network (PB.net) and the Earth
League's EarthDoc programme.

\end{document}